\documentclass[pss,fleqn]{w-art}
\usepackage{times}
\usepackage{w-thm}
\theoremstyle{plain}

\theoremstyle{definition}

\usepackage[]{graphicx}
\chardef\bslash=`\\ % p. 424, TeXbook

\begin{document}
\renewcommand{\copyrightyear}{2002}
\DOIsuffix{theDOIsuffix}
\Volume{XX}
\Issue{1}
\Month{01}
\Year{2006}
%%    First and last pagenumber of the article. If the option
%%    'autolastpage' is set (default) the second argument may be left empty.
\pagespan{1}{}
\Receiveddate{23 August 2002}
\Accepteddate{13 September 2002}

%% \pretitle{Editor's Choice}

\title[Unsual Quantum Phase in Exactly Solvable Ising-Heisenberg Models]{Unusual Quantum Phase in Exactly Solvable Doubly Decorated Ising-Heisenberg Models}

\author{Jozef Stre\v{c}ka\footnote{E-mail: {\sf jozkos@pobox.sk}, Phone: +421\,55\,6222121\,no.231,
           Fax: +421\,55\,6222124} and Michal Ja\v{s}\v{c}ur} 
\address[]{Department of Theoretical Physics and Astrophysics, 
Faculty of Science, \\ P. J. \v{S}af\'arik University,  
Park Angelinum 9, 040 01 Ko\v{s}ice, Slovak Republic}

\maketitle                  

The quantum Heisenberg model belongs to the most intensively studied models in statistical 
mechanics (see [1] and references therein), since it exhibits very interesting quantum phenomena.
However, due to the mathematical complexities related to the quantum nature of the model, 
only few exact results have been obtained in this field so far. Owing to this fact, the main aim 
of this work is to introduce a new statistical model, which is exactly solvable for planar lattices 
and simultaneously exhibits interesting quantum properties.
The model to be investigated here consists of Ising- and Heisenberg-type atoms regularly distributed
on planar doubly decorated lattices, as it is depicted in Fig. 1 for the case of the square
lattice. In order to derive the exact results for the model under investigation, it is very 
convenient to express the total Hamiltonian in the form 
$\hat {\cal H}_d = \sum_{k=1}^{Nq/2} \hat {\cal H}_k$, where the summation is carried out
over all the bonds of the original (undecorated) lattice, $N$ denotes the total number of the 
Ising-type atoms and $q$ is the coordination number of the original (undecorated) lattice. 
The bond Hamiltonian $\hat {\cal H}_k$ is given by (see Fig. 1)
\begin{eqnarray}
\hat {\mathcal H}_{k} = - J \left[\Delta (\hat S_{k1}^x \hat S_{k2}^x 
                      + \hat S_{k1}^y \hat S_{k2}^y) + \hat S_{k1}^z \hat S_{k2}^z \right]
                      - J_1 (\hat S_{k1}^z + \hat S_{k2}^z) (\hat \mu_{k1}^z + \hat \mu_{k2}^z)
                      - D \left[(\hat S_{k1}^z)^2 + (\hat S_{k2}^z)^2 \right],     
\label{HD}
\end{eqnarray}
where $\hat S_{k \alpha}^{\gamma}$ ($\gamma = x,y,z; \alpha = 1,2$) represent the relevant 
components of the standard spin-1 operators and $\mu_{k \alpha}^z$ those of the spin-1/2 operators. 
The exchange interaction $J_1$ couples the nearest-neighboring Ising and Heisenberg atoms and 
similarly, $J$ couples nearest-neighboring Heisenberg pairs. Finally, $\Delta$ describes the 
\begin{figure}[htb]
\vspace{0.5cm}
\begin{center}
\includegraphics[width=8cm]{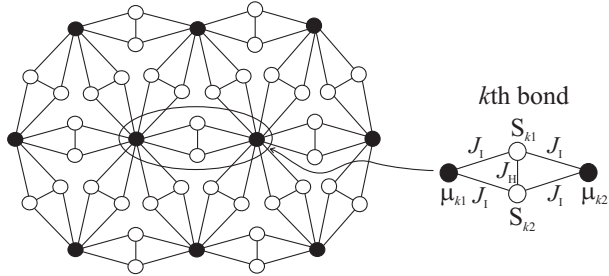}
\end{center}
\vspace{-0.2cm}
\caption{Part of the doubly decorated square lattice. Black circles represent the Ising atoms 
with spin 1/2 and gray ones denote the decorating Heisenberg atoms with spin 1.}
\label{fig1}
\end{figure}
anisotropy in the Heisenberg interactions and $D$ denotes the single-ion anisotropy. Taking into account the standard commutation relations for the bond Hamiltonians (i.e. $[\hat {\mathcal H}_{i}, \hat {\mathcal H}_{k}] = 0$, $i \neq k$), one can rewrite the partition function of the decorated model in the form
\begin{eqnarray}
{\mathcal Z}_{d} = \mbox{Tr} \exp (- \beta \hat {\cal H}_d ) 
                 = \mbox{Tr}_{\{ \mu \}} \prod_{k=1}^{Nq/2} \mbox{Tr}_{S_{k1}} \mbox{Tr}_{S_{k2}} 
                  \exp (- \beta \hat {\cal H}_k).     
\label{ZD}
\end{eqnarray}
Here, the symbol $\mbox{Tr}_{\{ \mu \}}$ means a trace over all degrees of freedom of Ising spins, $\mbox{Tr}_{S_{k1}} \mbox{Tr}_{S_{k2}}$ denotes a trace over a couple of Heisenberg spins residing on the $k$-th bond and $\beta = 1/(k_{\rm B} T)$. To proceed further, we introduce an extended decoration-iteration transformation [2]. Namely,
\begin{eqnarray}
\mbox{Tr}_{S_{k1}} \mbox{Tr}_{S_{k2}} \exp (- \beta \hat {\cal H}_k)
 = A \exp(\beta R \mu_{k1}^z \mu_{k2}^z),     
\label{MT}
\end{eqnarray}
from which, after evaluating the l.h.s., we obtain for the transformation parameters 
$A$ and $R$ the following expressions:
\begin{eqnarray}
A = \sqrt{(W_0 + W_1)(W_0 + W_2)}, \quad \beta R = 2 \ln(W_0 + W_1) - 2 \ln(W_0 + W_2),   
\label{MP}
\end{eqnarray}
where
\begin{eqnarray}
W_0 &=& \exp [\beta (2D-J)] + 2 \exp[\beta (D-J/2)] 
\cosh \left[ \frac{\beta}{2} \sqrt{(2D-J)^2 + 8 (J \Delta)^2} \right], \\
W_1 &=& 2 \exp [\beta (2D+J)] \cosh (2 \beta J_1) 
    + 4 \exp(\beta D) \cosh(\beta J \Delta) \cosh (\beta J_1), \\
W_2 &=& 2 \exp [\beta (2D+J)] + 4 \exp(\beta D) \cosh(\beta J \Delta).   
\label{MPD}
\end{eqnarray}
Finally, after substituting (3) into (2) one obtains the equation ${\cal Z}_d = A^{Nq/2} {\cal Z}_{0}$, which relates the partition function of doubly decorated Ising-Heisenberg model (${\cal Z}_d$) to that one of the original (undecorated) spin-1/2 Ising model (${\cal Z}_0$), that is exactly known for many planar lattices. Although, from this simple relation we can directly calculate all relevant physical quantities over the whole range of the parameters and temperature, we restrict ourselves in this brief report only to the analysis of the ground-state properties of the system. For this purpose, we have studied the lowtemperature behavior of the sublattice magnetization $m_A = \langle \hat \mu_{k \alpha}^z \rangle$, $m_B = \langle \hat S_{k \alpha}^z \rangle$, as well as the variations of the relevant nearest-neighbor correlation functions, namely, $q_{ii}^{zz} = \langle \hat \mu_{k1}^z \hat \mu_{k1}^z \rangle$, $q_{ih}^{zz} = \langle \hat \mu_{k1}^z \hat S_{k1}^z \rangle$, $q_{hh}^{\gamma \gamma} = \langle \hat S_{k1}^{\gamma} \hat S_{k2}^{\gamma} \rangle$ ($\gamma = x,y,z$) and 
$\eta = \langle (\hat \hat S_{k1}^z)^2 \rangle$. Before discussing the results, it is worth 
noticing that the ground state of our system is universal in the sense that it does not depend 
on the coordination number and the space dimensionality of the lattice.

At first, we have depicted in Fig. 2a the ground-state phase diagram in the $D-J1$ space for the
system without single-ion anisotropy, i.e. $D/J = 0.0$. As one can see, three different phases 
can appear in the system depending on the ratio between $J_1$ and $D$. The phase denoted as O$_1$ represents a standard ferromagnetic phase which is usually observed in the pure Ising systems. 
As one can expect, in this case we have $m_A = 0.5$, $m_B = 1.0$, $q_{ii}^{zz} = 0.25$, $q_{ih}^{zz} = 0.5$, $q_{hh}^{zz} = 1.0$, $\eta = 1.0$ and $q_{hh}^{xx} = q_{hh}^{yy} = 0.0$. On the other hand, 
the phase denoted as O$_2$ exhibits some unusual features, though it is also a long-range ordered ferromagnetic phase in which all Ising spins are perfectly aligned along the easy axis (thus, 
$m_A = 0.5$ and $q_{ii}^{zz} = 0.25$). Contrary to the perfectly ordered Ising sublattice, the Heisenberg one behaves in a different way. Indeed one finds that $m_B = 0.5$, $q_{ih}^{zz} = 0.25$, $q_{hh}^{zz} = 0.0$ and $\eta = 0.5$. Moreover, in this phase one also observes the nonvanishing
short-range order in the $xy$ plane ($q_{hh}^{xx} = q_{hh}^{yy} = 0.5$). This rather unexpected 
\begin{figure}[htb]
\vspace{-0.1cm}
\begin{minipage}[t]{0.48\textwidth}
\includegraphics[width=1\textwidth]{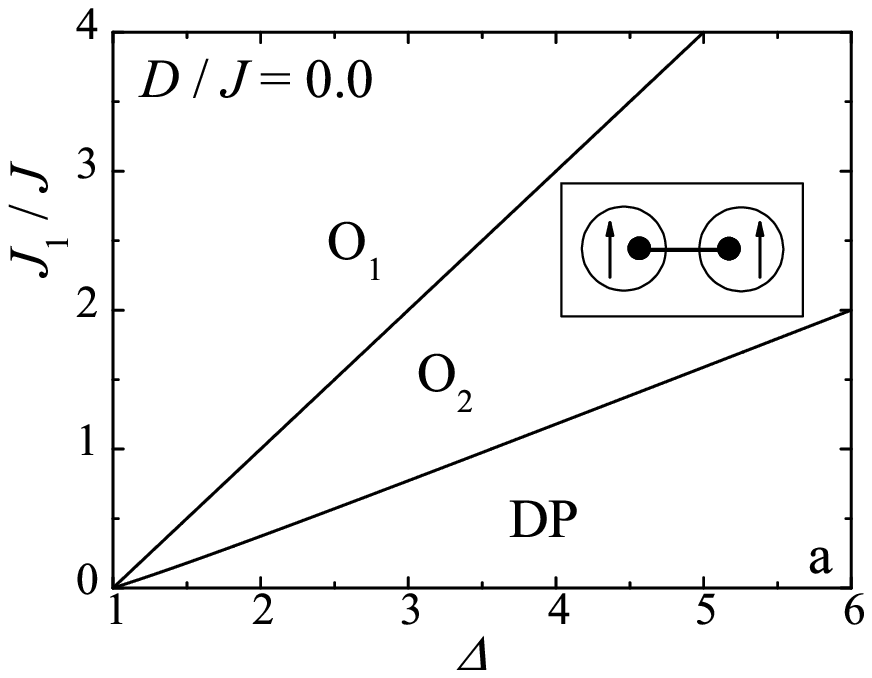}
\end{minipage}
\vspace*{0.5cm}
\begin{minipage}[t]{0.48\textwidth}
\includegraphics[width=0.96\textwidth]{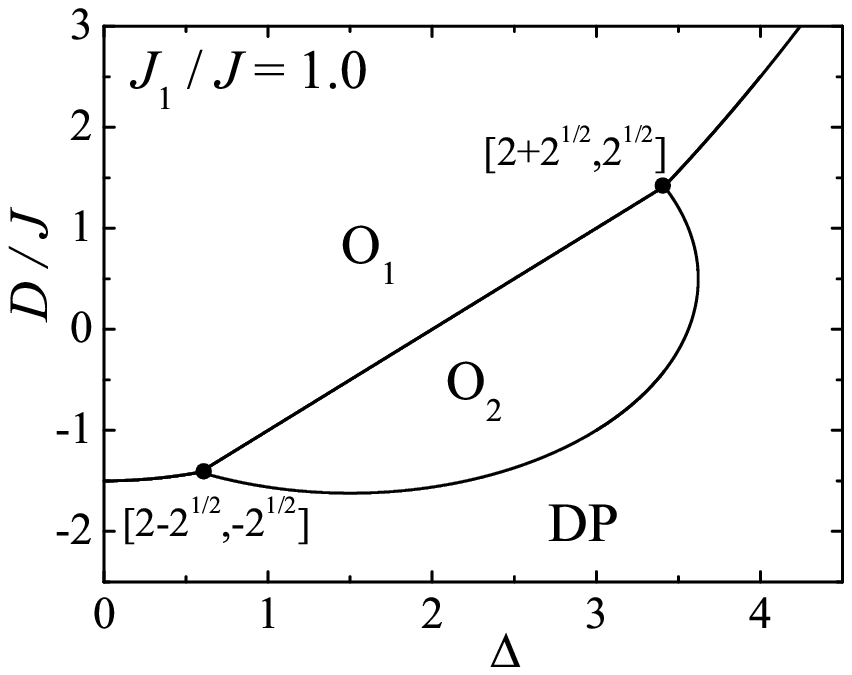}
\end{minipage}
\vspace*{-0.8cm}
\caption{Ground-state phase diagrams of the doubly decorated Ising-Heisenberg model. O$_1$ means the
standard ferromagnetic phase, O$_2$ represents the quantum ferromagnetic phase, the properties of which
can be explained within the valence-bond-solid picture. The situation is schematically depicted for the
decorating atoms in the inset of Fig. 2a. The large open circles represent the atomic sites with $S=1$, the small solid circles and arrows denote the spin 1/2 variables. The solid line shows the valence bond. Finally, DP denotes the disordered frustrated phase and the black points mark in Fig. 2b the triple points.}
\label{fig2}
\end{figure}
behavior is simply explainable within the valence-bond-solid picture [3], that is schematically depicted for the decorating atoms in the inset of Fig. 2a. This is, of course, the pure quantum 
effect which is caused by relatively strong quantum fluctuations in the system. It is also clear 
that the phase O$_2$ represents a mixture of regularly ordered Ising spins with the quantum phase 
on the Heisenberg sublattice and to the best of our knowledge such a phase has not been discussed 
in the literature before. Finally, if the anisotropy parameter $\Delta$ becomes strong enough, 
a disordered phase (DP) appears in the system. In addition to the trivial fact that in this phase 
$m_A = m_B = 0.0$ is found, one also finds $q_{ii}^{zz} = 0.25$, $q_{ih}^{zz} = q_{ih}^{zz} = 0.0$,
$q_{hh}^{zz} < 0.0$ and $q_{hh}^{xx} = q_{hh}^{yy} > 0.0$. In fact, all these findings clearly 
indicate that the relevant disordered quantum phase is strongly frustrated. It is also worth 
noticing that the boundary between the phase O$_1$ and O$_2$ is given by $J_1 = J (\Delta -1)$ 
and that separating the phase O$_2$ and DP is given by $J_1 = J (0.5 \sqrt{1 + 8 \Delta^2} 
- \Delta - 0.5)$.

To complete the ground-state analysis, we have shown in Fig. 2b the phase diagram of the model
in the $\Delta - D$ space for $J_1 = J$. Similarly, as in the previous case we have again observed the
same phases. The phase diagram, however, becomes more interesting, since in this case we have
also found two triple points, namely, T$_1$ = $[2 - \sqrt{2}, -\sqrt{2}]$ and T$_2$ = $[2 + \sqrt{2}, \sqrt{2}]$. Moreover, it is apparent that the relevant phase boundaries become more complex. 
Here we only state that in both cases ($D = 0$ and $D \neq 0$) all phase boundaries represent the points at which the system exhibits first-order phase transitions. The detailed analysis of the 
ground state properties, as well as that one of the thermal behavior of the system is being left 
for our future work which is now in preparation.

To conclude this work, we would like to emphasize that the Ising-Heisenberg models (including
the present model) represent very useful exactly solvable statistical models that can exhibit very
interesting quantum phenomena. Moreover, these models can be helpful in description of some
real materials. For example, the recently synthesized compound studied in Ref. [4] has the structure
of the doubly decorated honeycomb lattice.

\end{document}